\documentclass[12pt,nohyper]{JHEP3}         
\usepackage{amssymb,amsfonts}

\newcommand{\be}{\begin{equation}}
\newcommand{\ee}{\end{equation}}
\newcommand{\bea}{\begin{eqnarray}}
\newcommand{\eea}{\end{eqnarray}}
\newcommand{\bref}[1]{(\ref{#1})}
\newcommand{\nn}{\nonumber}
  
 \newcommand{\D}{\delta} 
\newcommand{\ep}{\epsilon} \newcommand{\vep}{\varepsilon}
\newcommand{\T}{\theta}

          \newcommand{\w}{\omega}
\newcommand{\z}{\zeta}          
\newcommand{\h}{\eta}

\newcommand{\cg}{\hat g}
\newcommand{\cga}{\hat \gamma}

\newcommand{\Tb}{{\overline\theta}}

\def\6{\partial}
\def\7{\tilde}
\def\8{\widehat}
 
\def\vs{\vskip 3mm}

\def\pa{\partial}

\def\CL{{\cal L}}

\def\bP{{\rm{\bf P}}}

\def\G11{\Gamma_{11} }

\def\lag{lagrangian }

\title{Non-Relativistic Superbranes}

\author{Joaquim Gomis\\
         Departament ECM, Facultat de F{\'\i}sica, \\
Universitat de Barcelona and Institut de F{\'\i}sica d'Altes Energies, \\
Diagonal 647, E-08028 Barcelona, Spain \\
        Email: \email{gomis@ecm.ub.es }}

\author{Kiyoshi Kamimura\\
        Department of Physics, Toho University\\ 
        Funabashi\ 274-8510, Japan\\
        Email: \email{kamimura@ph.sci.toho-u.ac.jp}}

\author{Paul K. Townsend\\ 
Department of Applied Mathematics and Theoretical Physics \\
Centre for Mathematical Sciences, University of Cambridge\\
Wilberforce Road, Cambridge, CB3 0WA, UK.\\
 Email: \email{p.k.townsend@damtp.cam.ac.uk}}

\abstract{Subtleties arising in the non-relativistic limit of 
relativistic branes are resolved, and a reparametrization-invariant 
and kappa-symmetric non-relativistic super p-brane action is obtained
as a limit of the action for a relativistic 
super p-brane in a Minkowski vacuum. We give explicit results 
for the D0-brane, which provides a realization of the 
super-Bargmann algebra, the IIA superstring and the 11-dimensional 
supermembrane.} 
\preprint{UB-ECM-PF-04/26, Toho-CP-0476, DAMTP-2004-94}

\begin{document}

\section{Introduction}\label{sec.1}

\noindent

It is often useful to consider special limits of physical theories 
and the  non-relativistic limit of relativistic theories is a good 
example. Non-relativistic particle mechanics is of course a venerable 
subject and so, in a sense, is the non-relativistic mechanics of branes; 
we have in mind the textbook studies of waves on strings and membranes, 
but since these involve some material medium their dynamics is not 
usefully thought of as the non-relativistic limit of some relativistic 
system. Alternatively, one can start with a relativistic p-brane and 
take its non-relativistic limit. This is standard  for $p=0$, although 
there are some well-known subtleties, but a new difficulty occurs for 
$p>0$. In these cases, each p-volume element, or `particle', of a 
closed p-brane has its `anti-particle'   elsewhere on the brane. 
As the characteristic feature of a non-relativistic theory is the 
absence of anti-particles, any non-relativistic limit of a p-brane 
must decouple `opposite'  p-volume elements. One can achieve this 
decoupling by rescaling  the spacetime coordinates in order  to 
focus on a small region of the brane and then taking a limit that 
decouples it from other regions. Let us illustrate this procedure 
with the action
\be
S= -T_p \int d^{p+1}\xi \, \sqrt{-\det G}
\ee
for a bosonic p-brane in a $d$-dimensional product spacetime, with metric
\be
ds^2 = \omega^2 \eta_{\mu\nu} dx^\mu dx^\nu + G_{ab}(X) dX^adX^b
\ee
where $\eta_{\mu\nu}= {\rm diag}(-1,1,\dots,1)$ is the Minkowski 
(p+1)-metric  (in cartesian coordinates), and $G_{ab}$ is a metric 
on some $(d-p-1)$-dimensional Riemannian  manifold ${\cal M}$. We 
have rescaled the coordinates of the Minkowski factor by introducing 
an arbitrary dimensionless constant $\omega$. The induced worldvolume metric is
\be
G_{ij} = \omega^2 g_{ij} + \partial_i X^a  \partial_j X^b G_{ab}
\ee
where 
\be
g_{ij} = \partial_i x^\mu \partial_j x^\nu\eta_{\mu\nu}
\ee
is the Minkowski metric in the arbitrary coordinates $\xi^i(x)$. It 
follows that
\be
\sqrt{-\det G} = \omega^{p+1} \sqrt{-\det g}\left[1+ {1\over2\omega^2} 
g^{ij}\partial_i  X^a  \partial_j X^b G_{ab} + {\cal O}
\left(1/\omega^4 \right)\right].
\ee
Defining a rescaled p-volume tension $T$ by
\be
T_p= \omega^{1-p}\, T, 
\ee
we find that
\be\label{lowenergy}
S = -T \int d^{p+1}\xi \left[ \omega^2\, \sqrt{-\det g}\ +\  
{1\over2} \sqrt{-\det g}\, g^{ij} \partial_i X^a \partial_j X^b G_{ab}  
\ + \dots \right], 
\ee
where the omitted terms vanish in the $\omega\rightarrow\infty$ limit, 
and are higher-order in derivatives. In effect, the expansion in  
inverse powers of $\omega$ is  an expansion in derivatives,  so the 
leading terms provide an effective low-energy action. This is 
generally what is meant by the `field theory limit', which therefore 
has an interpretation as a non-relativistic limit \cite{Townsend:1999hi,
Gomis:2000bd, Danielsson:2000gi, Garcia:2002fa}.

Note now that  $\sqrt{-\det g}$ is the Jacobian for the change 
of Minkowski space coordinates from the cartesian coordinates $x^\mu$ 
to the arbitrary coordinates $\xi^i$, so the first term  in 
(\ref{lowenergy}) is the integral over the Minkowski spacetime 
of a constant proportional to $\omega^2T$.  Alternatively, one may observe 
that 
$\sqrt{-\det g}$ is a total derivative with respect to the $\xi$ coordinates.
From either perspective, it is clear that this term may be omitted. 
For $p=0$ this corresponds to the subtraction of the rest-mass 
energy from the total energy. For $p=1$ it corresponds to the 
subtraction procedure advocated in \cite{Gomis:2000bd, Danielsson:2000gi}. 
Having removed this term we can take the $\omega\rightarrow\infty$ 
limit to arrive at the 
(still reparametrization invariant) {\it non-relativistic} p-brane action
\be\label{reparamp}
S =  - {1\over2}T  \int d^{p+1}\xi \sqrt{-\det g}\, g^{ij} 
\partial_i X^a \partial_j X^b G_{ab}. 
\ee
For $p=0$, we can set $T=mc$ and $x^0(\xi^0)= ct(\xi^0)$ to get 
the standard time-reparametrization invariant action for a 
non-relativistic particle on ${\cal M}$:
\be\label{pzerocase}
S_0 = m \int d\xi^0 \, {|\dot X|^2 \over 2 \dot t} \, .
\ee
The action (\ref{reparamp})  generalizes this to $p>0$. 

If we choose the Monge gauge
\be
\partial_i x^\mu= \delta_i{}^\mu
\ee
to fix the reparametrization invariance of (\ref{reparamp}) 
then we get the standard Minkowski space sigma-model action 
\be
S= -{1\over2} T \int d^{p+1}\xi \, \eta^{ij}\partial_i X^a 
\partial_j X^b G_{ab}(X). 
\ee
Note that disturbances of the sigma-model  fields  propagate at the 
speed of light, which we could have set to unity from the beginning. 
This may seem puzzling, because  one is used to thinking of the 
non-relativistic limit as one in which $c\rightarrow\infty$, but 
$c$ is not dimensionless so, strictly speaking, it makes no sense 
to think of it as a variable that we can take to infinity \cite{Duff:2002vp}. 
For particles, one can get away with thinking of the non-relativistic 
limit as one in 
which $c\rightarrow\infty$. However, it is preferable not to 
think of it this way, and essential not to do so for branes. 

The purpose of this paper is to generalize some of the above 
discussion to super-p-branes. The most general action that we will consider  
here takes the form
\be
S= -T_p \int d^{p+1}\xi \, \sqrt{-\det G} + Q_p \int \! b
\ee
where the worldvolume metric $G$ is a superspace extension of 
the induced worldvolume metric and $b$ is a superspace
$(p+1)$-form constructed from a super-Poincar\'e invariant 
closed  $(p+2)$-form $h=db$. For the moment, we allow this
`Wess-Zumino' (WZ) term to have 
an arbitrary coefficient $Q_p$, which can be interpreted as the 
p-brane charge. We will suppose that the superspace  background 
is the Minkowski vacuum of some supergravity theory, and that the
action is invariant under the action of the super-Poincar\'e isometry 
group of this background. In particular, we assume invariance under
the infinitesimal supersymmetry transformation  
\be\label{susy}
\delta \theta = \epsilon \, , \qquad \delta X^m =  
-i\bar\epsilon \Gamma^m \theta. 
\ee
We suppose, for the moment,  that $\theta$ is a minimal spinor. 
For simplicity of presentation we also assume  that the spacetime 
dimension $d$ is such that the Dirac matrices $\Gamma^m$ 
($m=0,1,\dots,d-1$) can be chosen to be real, in which case 
$\theta$ is real, and 
\be
\bar\T = \T^T\Gamma^0. 
\ee
In other words, we assume that $\T$ is Majorana (or Majorana-Weyl) and choose a (real) basis of the 
Dirac matrices for which the charge conjugation matrix $C$ is equal to 
$\Gamma^0$.

Both the induced metric $G_{ij}$ and the superspace $(p+2)$-form 
$h$ are constructed from the super-translation invariant  1-forms
$d\T$ and 
\be
\Pi^m = dX^m + i \bar\theta\Gamma^m d\theta, 
\ee
Specifically,
\be
G_{ij} = \Pi_i{}^m \Pi_j{}^m \eta_{mn} , 
\ee
where $\Pi_i{}^m$ are the components of the worldvolume 1-forms induced by $\Pi^m$, and 
\be
h= -{i \over p!}\; \Pi^{m_1}\wedge \dots \wedge \Pi^{m_p} d\bar\theta 
\Gamma_{m_1\dots m_p}d \theta.
\ee
We must assume that $h$ does not vanish identically. The requirement that $dh=0$ is then 
equivalent to \cite{Bergshoeff:1987cm}
\be\label{matrixidentity}
\left(\bar u \Gamma^m u\right)\left(\bar u\Gamma_{mn_1\dots n_p} u\right) \equiv 0
\ee
for {\it commuting} minimal spinor $u$. Given our restriction to spacetime dimensions for 
which the Dirac matrices are real, this restricts the worldspace dimension to 
$p=1,2,\,{\rm mod}\, 4$. This does not include $p=0$, but we will later 
consider separately the D0-brane (for which the spinor $\T$ is not minimal). 

Suppose now that we rescale the superspace coordinates by setting
\be\label{rescale}
X^m = (\omega x^\mu, X^a) , \qquad \theta = 
{1\over \sqrt{\omega}}\, \vartheta. 
\ee
The induced metric is then
\be
G_{ij} = \omega^2\left[ g_{ij} + {1\over \omega^2}\,  
\partial_i {\bf X} \cdot \partial_j {\bf X} + {2 \over\w^2}\, i
\bar\vartheta \gamma_{(i}\partial_{j)} \vartheta + 
{\cal O}\left(1/\w^4\right)\right]
\ee
where
\be
\gamma_i = \partial_i x^\mu \Gamma_\mu \, 
\ee
and hence 
\be
\sqrt{-\det G} =  \omega^{p-1}\, \sqrt{-\det g}\left[ \omega^2 + 
{1\over2} g^{ij}\partial_i {\bf X}\cdot \partial_j {\bf X} + 
i \bar\vartheta \gamma^i\partial_i \vartheta  + 
{\cal O}\left(1/\w^2\right)\right] 
\ee
where $\gamma^i= g^{ij}\gamma_j$. Similarly, 
\be
h= -{i\over p!}\, \omega^{p-1}\left[ dx^{\mu_1}\wedge \dots \wedge dx^{\mu_p} 
d\bar\vartheta \Gamma_{\mu_1\dots \mu_p}d\vartheta + 
{\cal O}\left(1/\w^2\right)\right],
\ee
which leads to a $(p+1)$-form  $b$ that is the worldvolume Poincar\'e dual 
of the WZ  scalar density
\be
{\cal L}_{WZ} =\w^{p-1}\left[\sqrt{-\det g}\;i(\bar\vartheta_- 
\gamma^i\partial_i 
\vartheta_--  \bar\vartheta_+ \gamma^i\partial_i 
\vartheta_+) +  {\cal O}\left(1/\w^2\right)\right].
\ee
Note that we now have an expansion `in derivatives'  for which a 
fermion counts as `half a derivative', as is generally the case 
for low-energy expansions, so the limit as $\omega\rightarrow\infty$ 
can again be viewed as a `field theory' limit.  If we now define a 
rescaled tension $T$ and rescaled charge $Q$ by
\be
T_p= \omega^{1-p}\, T,  \qquad Q_p= \omega^{1-p}\, Q\, , 
\ee
and follow the same procedure as before, then we arrive at the 
free-field theory with 
Lagrangian density
\be\label{freefield}
{\cal L} = -{T\over 2} \eta^{ij}  \partial_i  {\bf X} \cdot \partial_j {\bf X}
 - {i} \left(T+Q\right)\bar\vartheta_+ \gamma^i \partial_i \vartheta_+ 
 - {i} \left(T-Q\right)\bar\vartheta_-\gamma^i \partial_i \vartheta_- 
 \ee
 where $\vartheta_\pm$ are eigenspinors with eigenvalue $\pm1$ of the matrix
\be\label{pstar}
\Gamma_* = {1\over \left(p+1\right)!}\, \varepsilon^{\mu_1\dots \mu_{p+1}} 
\Gamma_{\mu_1\dots \mu_{p+1}}, 
\ee
which satisfies $\Gamma_*^2=1$ as a consequence of the previously 
mentioned restriction to  $p=1,2$ mod $4$.  Note that $T<|Q|$ implies 
ghosts in the quantum theory, so that $T$ should satisfy the bound 
$T\ge |Q|$. 

So far, our extension of the non-relativistic limit of bosonic 
p-branes to super-p-branes appears satisfactory, and it is for 
$T> |Q|$. Before the rescaling of coordinates, the action was invariant 
under the 
rigid supersymmetry transformations (\ref{susy}). For these 
transformations  to have a sensible $\omega\rightarrow\infty$ 
limit after the rescaling, we must define a rescaled supersymmetry 
parameter $\varepsilon= \sqrt{\omega}\,  \epsilon$. Then, in the 
$\omega\rightarrow\infty$ limit, the only non-zero `supersymmetry' transformation is 
\be
\delta\vartheta = \varepsilon. 
\ee
This is certainly a symmetry of the non-relativistic action, and is all 
one should expect for $T>|Q|$; it is the linearized remnant of a 
non-linear realization of spacetime
supersymmetry. However, the $T=|Q|$ case is 
clearly special:  in this case we know that the relativistic action has a
fermionic gauge invariance, called `kappa-symmetry', that allows half 
the fermions to be `gauged away', so it should not be a surprise that 
half of the fermions drop out in the 
non-relativistic limit.
However, we also know that when $T=|Q|$ the 
gauge-fixed relativistic action has a linearly-realized 
worldvolume supersymmetry
\cite{Hughes:1986dn, Achucarro:1989ev}. We would expect this feature to survive 
the non-relativistic limit, and it is obvious that the free field Lagrangian (\ref{freefield}) 
{\it is} supersymmetric when $T=|Q|$, because of the match of bose/fermi degrees of freedom 
implied by the assumed restrictions on $(d,p)$ \cite{Achucarro:1987nc}. However, {\it this 
linearly-realized worldvolume supersymmetry of the non-relativistic super-p-brane 
action appears not to be the limit of any symmetry of the relativistic action}. 

The resolution of the puzzle is that precisely when $T=|Q|$, but not 
otherwise, there is {\it another} non-relativistic limit, differing 
from the one just described in the way that the fermionic variables 
are rescaled. Instead of (\ref{rescale}) we may set
\be\label{newrescale}
X^m = (\omega x^\mu, X^a) , \qquad 
\theta = \sqrt{\omega}\; \theta_- + {1\over\sqrt{\omega}}\, \theta_+
\ee
where $\theta_\pm$ are $\pm1$ eigenspinors of  $\Gamma_*$.  Because 
$\T_-$ is scaled by a {\it positive} power 
of $\omega$, the action for general $T,Q$ has no $\omega\rightarrow\infty$ 
limit. However, a cancellation occurs for $T=|Q|$ that makes the limit possible
(once a constant  `rest-energy'  term has been discarded). Note that the 
new limit is {\it not} one in which all  fermions count as `half a 
derivative'; this is true only of $\T_+$. However, we will see that  
kappa-symmetry will allow us to set $\T_-=0$, so that the new limit is
still a field theory limit in terms of the physical worldvolume fields. For the
Minkowski background considered here, both `old'  and `new' non-relativistic limits 
yield the same final (free-field) action, {\it but only in the new limit is
the linearly realized supersymmetry  of this action obtained 
as a limit of the symmetries of the relativistic action}.

We will begin with  a discussion of the non-relativistic limit of the 
kappa-symmetric super-p-brane action, along the lines indicated above. 
We will show that this limit results in a non-relativistic super-p-brane
action that is both invariant under a non-relativistic spacetime supersymmetry, and 
gauge invariant with respect to a non-relativistic kappa-symmetry. 
This action is given implicitly. for general $p$, in terms of 
a `non-relativistic' closed superspace 
$(p+2)$-form $h_{nr}$, but we give the explicit result 
for $p=2$, which includes the case of the 11-dimensional supermembrane 
(M2-brane). Moreover,  the gauge fixed action is easily found
 for general (admissible) $p$, and the result is shown to be a 
$(p+1)$-dimensional 
super-Poincar\'e invariant worldvolume field theory. This transmutation of spacetime into 
worldvolume supersymmetry  occurs also for the relativistic super-p-brane, 
but it is much easier to see in the non-relativistic limit. 

We then move to a discussion of the non-relativistic limit of the 
D0-brane. This is properly thought of as the $p=0$ subcase of the super 
D-p-brane rather than as the $p=0$ subcase of the super-p-brane that we 
have been describing above, so there are a few differences. However, the
basic idea is the same, and this case has the advantage that 
the explicit gauge-invariant non-relativistic action is simple. 
For this case, we also consider
the effect of the non-relativistic limit on the super-Poincar\'e 
symmetry algebra of the 
D0-brane.  One of the subtleties of the non-relativistic limit 
alluded to earlier is that the
symmetry algebra of the non-relativistic bosonic particle is 
{\it not} the Galilei algebra obtained by contraction of the 
Poincar\'e symmetry algebra of the relativistic particle, but 
a central extension of it called the Bargmann algebra. The 
analogous central charges for $p>0$ were discussed in a recent 
article \cite{Brugues:2004an}. For the D0-brane, however, the 
super-Bargmann symmetry 
algebra of the non-relativistic D0-brane {\it is} the 
contraction of the super-Poincar\'e symmetry algebra of 
the relativistic D0-brane, and we show explicitly how it  
is realized. 

Finally, we consider the IIA superstring, presenting an explicit 
reparametrization invariant and kappa-symmetric non-relativistic 
action, and conclude with some comments on future directions

\vs

\section{The non-relativistic super-p-brane}
\label{sec.2}

We now concentrate on the kappa-symmetric super-p-brane action
\be
S= -T_p \int d^{p+1}\xi \, \left[ \sqrt{-\det G} - {\cal L}_{WZ}\right]
\ee
where ${\cal L}_{WZ}$ is the Wess-Zumino Lagrangian density. This action is 
invariant under the `kappa'-gauge transformation 
\be\label{kapsym}
\delta\T = {1\over2}\left(1- \Gamma_\kappa\right)\kappa(\xi), \qquad 
\delta X^m = -i\bar\T\Gamma^m\delta\T\, , 
\ee
where \cite{Hughes:1986fa,Bergshoeff:1987cm}
\be
\sqrt{-\det G}\ \Gamma_\kappa= {1\over (p+1)!}\, 
\varepsilon^{i_1\dots i_{p+1}} \Pi_{i_1}^{m_1} \cdots \Pi_{i_{p+1}}^{m_{p+1}} 
\Gamma_{m_1\dots m_{p+1}}. 
\ee
This matrix satisfies $\Gamma_\kappa^2=1$ (assuming the 
restrictions spelled out earlier). 

To take the non-relativistic limit, we first define the 1-forms
\be
e^\mu ~= dx^\mu +{i}\Tb_-\Gamma^\mu d\T_-,
\qquad u^a  ~=~dx^a+{2i}\Tb_+\Gamma^a d\T_-\ , 
\ee
where (for later convenience) we have introduced the new transverse space coordinate
\be\label{xX}
x^a ~= ~X^a+{i}\Tb_-\Gamma^a\T_+. 
\ee
We then define rescaled coordinates as in  (\ref{newrescale}), so that
\be
\Pi_i^\mu = \omega e_i{}^\mu + {i\over\omega} 
\bar\theta_+\Gamma^\mu \pa_i\theta_+, \qquad
\Pi_i{}^a = u_i{}^a.
\ee
This yields
\be
G_{ij} = \omega^2\, \cg_{ij} + 2i \bar\theta_+\cga_{(i} \partial_{j)} 
\theta_+ + {\bf u}_i \cdot {\bf u}_j + {\cal O}\left(1/\omega^2\right) 
\ee
where\footnote{The hats are to remind us that these quantities depend on $\T_-$.}
\be
\cg_{ij} = e_i{}^\mu e_j{}^\nu\eta_{\mu\nu},  
\qquad \cga_i = e_i{}^\mu\Gamma_\mu. 
\ee
Noting that
\be
\sqrt{-\det \cg} = \det e_i{}^\mu  =:  e
\ee
we now find that
\be\label{NGexpansion}
\sqrt{-\det G} = \omega^{p-1}\, e \left[ \omega^2\, + 
\,{1\over2} \cg^{ij}{\bf u}_i \cdot {\bf u}_j + 
i\bar\theta_+\cga^i \partial_i\theta_+  + {\cal O}\left(1/\w^2\right) \right] , 
\ee
where $\cg^{ij}$ is the inverse of $\cg_{ij}$ and 
$\cga^i = \cg^{ij}\cga_j$. 

To complete the expansion of the action in inverse powers of $\omega$, 
we need the expansion of the superspace $(p+2)$-form $h$. With the exterior 
product of forms being understood, this is
\be
h = \omega^{p+1}h_0 + \omega^{p-1}\, h_{nr} + {\cal O}(\w^{p-3})
\ee
where
\be
h_0 = -{i\over p!}\left(d\bar\T_-\hat\gamma_{i_1\dots i_p} 
d \T_-\right)d\xi^{i_1} \cdots  d\xi^{i_p}
\ee
and 
\bea
h_{nr}&=&-{i \over p!} \bigg[
\left(d\Tb_+\hat\gamma_{i_1\dots i_{p}}d\T_+\right)~+~
\nn\\&&
+p 
\left(d\bar\T_- \cga_{i_1\dots i_{p-1}}\Gamma_\mu d\T_-\right) 
\left(i\Tb_+\Gamma^\mu\partial_{i_p}\T_+\right)
+ \ 2 p\, \left(d\Tb_+\cga_{i_1\dots i_{p-1}} 
\Gamma_ad\T_-\right)u_{i_p}{}^a  \nonumber\\
&& +\ {p(p-1)\over2}\left(d\Tb_-\cga_{i_i\dots i_{p-2}}\Gamma_{ab}d\T_- \right)
u_{i_{p-1}}{}^a u_{i_p}{}^b
\bigg]d\xi^{i_1} \cdots  d\xi^{i_p}.
\eea
The leading term $h_0$ is not manifestly a closed form 
because $d\cga^i$ is non-zero. However, one can show by means of the
 identity \bref{matrixidentity}  and a 
lengthy algebraic manipulation (which we omit)
that $h_0 = db_0$, where
\be
b_0 = \left(d\xi^0\dots d\xi^p\right)\left[\sqrt{-\hat g} - \sqrt{-g}\right]. 
\ee
Similarly, it follows from the closure of $h$ that $h_{nr}$ is closed, 
so we can write (at least locally) 
\be
h_{nr} = db_{nr}
\ee
for some $(p+1)$-form $b_{nr}$; its explicit form is not easy to determine for general $p$, so we will limit ourselves here to giving it for $p=2$. 
\bea
b_{nr} &=& -{1\over2}\Bigg\{K^+_{\mu\nu}\left[e^\mu e^\nu - 
K_-^\mu e^\nu + {1\over3}K_-^\mu
K_-^\nu\right]
+  K^-_{\mu\nu}K_+^\mu \left[e^\nu - {1\over3}K_-^\nu\right]\nn\\
&& + \ L_{\mu b}\left[2e^\mu u^b - e^\mu L^b - K_-^\mu u^b + 
{2\over3}K_-^\mu L^b\right] \nn\\
&&+ \ K^-_{ab}\left[u^a u^b - L^a u^b + {1\over3}L^a L^b\right]\Bigg\}
\eea
where
\bea
K^\pm_{\mu\nu}& =& i \bar\T_\pm \Gamma_{\mu\nu} d\T_\pm, \qquad
K^\mu_\pm = i \bar\T_\pm \Gamma^\mu d\T_\pm, \qquad
K^-_{ab} = i \bar\T_- \Gamma_{ab} d\T_- , \nn\\
L_{\mu a} &=& i \bar\T_+ \Gamma_\mu \Gamma_a d\T_- +
i \bar\T_- \Gamma_\mu \Gamma_a d\T_+ , \qquad
L^a = i \bar\T_+ \Gamma^a d\T_- +i \bar\T_- \Gamma^a d\T_+. 
\eea 
This case is of particular interest as it includes the 11-dimensional supermembrane 
\cite{Bergshoeff:1987cm}. 

We are now in a position to compute the non-relativistic limit of the action.
In terms of the rescaled tension $T$, the total Lagrangian density is
\be
-T_p\sqrt{-\det G}+ T_p{\cal L}_{WZ} = -\w^2 T\det (\pa_i x^\mu) + 
T{\cal L}_{nr} + {\cal O}(\w^{-2}).
\ee
Because of a crucial cancellation, the term that is singular as 
$\w\rightarrow\infty$ 
is exactly the same  total derivative as in the bosonic case, and may be 
discarded. The $\w\rightarrow\infty$ limit may then be taken, leading to the 
non-relativistic action 
\be\label{nract}
S_{nr} = -T\int d^{p+1}\xi \, 
\sqrt{-\det g}\left[{1\over2} \cg^{ij}{\bf u}_i \cdot {\bf u}_j + 
i\bar\theta_+\cga^i \partial_i\theta_+\right] + T\int\! b_{nr}
\ee
As we shall see shortly, this Lagrangian simplifies dramatically after fixing 
the kappa gauge symmetry. First, we note that it is invariant under a
 `non-relativistic spacetime supersymmetry' . If one sets
\be
\epsilon = \sqrt{\w}\, \epsilon_- + {1\over\sqrt{\w}}\, \epsilon_+.
\ee
for (rescaled) $\Gamma_*$-eigenspinor parameters $\ep_\pm$, then the
the original supersymmetry transformations (\ref{susy}) have the $\w\rightarrow\infty$ limit
\be\label{nrsusy}
\delta \T_\pm = \ep_\pm, \qquad \delta x^\mu = i\bar\T_-\Gamma^\mu\ep_-, 
\qquad 
\delta x^a = 2i\bar\T_-\Gamma^a\ep_+. 
\ee
Note the simple transformation law of  $x^a$, defined in (\ref{xX}). 

The non-relativistic kappa-symmetry transformations are found in a similar 
way from the expansion
\be\label{gamexp}
\Gamma_\kappa= \Gamma_* - {1 \over \omega }
\left[ \cga^k u_k{}^a\Gamma_a  \Gamma_* 
\right]  + {\cal O}\left(\omega^{-2}\right).
\ee
In terms of the $\Gamma_*$ eigenspinor parameters $\kappa_\pm$ defined 
by 
\be
\kappa = \sqrt{\omega}\, \kappa_- + {1\over\sqrt{\omega}}\, \kappa_+, 
\ee
the original kappa-symmetry  transformations (\ref{kapsym}) become, 
in the $\w\rightarrow\infty$ limit,
\bea\label{nrk}
\delta \T_- &=&  \kappa_-  , 
\qquad \qquad \delta\T_+ = 
-{1\over2} \cga^i u_i{}^a\Gamma_a \kappa_-\nonumber\\
\delta x^\mu &=& -i\bar\T_-\Gamma^\mu \kappa_-, \qquad 
\delta x^a = -2i\bar\T_+\Gamma^a \kappa_-.
\eea
Note that only $\kappa_-$ appears in these transformations. 
As a consequence,
{\it  the non-relativistic kappa transformation is an irreducible 
gauge symmetry}, in contrast to the infinitely reducible gauge 
symmetry in the relativistic case. 

The non-relativistic gauge invariance (\ref{nrk}) allows us 
to set $\T_-=0$, in which case
\be
b_{nr} =-{i\over p!} (\T_+ \gamma_{i_1\dots i_p}
d\T_+)d\xi^{i_1}\cdots d\xi^{i_p}.
\ee
The non-relativistic Lagrangian density in this gauge is 
\be
{\cal L}_{nr} = - \sqrt{-\det g}\, \left[ {1\over2}g^{ij} 
 \partial_i  
{\bf x} \cdot \partial_j {\bf x}+
2i\bar\T_+ \gamma^i\partial_i\T_+\right]. 
\ee
Fixing the reparametrization invariance by the Monge gauge 
choice we arrive at the fully gauge-fixed Lagrangian density
\be
{\cal L} = -{1\over 2} \eta^{ij}  \partial_i  
{\bf x} \cdot \partial_j {\bf x}
 - 2i \theta_+ \Gamma^i \partial_i \theta_+ . 
\ee

Our non-relativistic limit was designed to capture the linearly realized 
supersymmetry as a limit of the (super)symmetries of the relativistic 
theory, so we should now verify that this has been accomplished. In the gauge 
$\T_-=0$, a supersymmetry transformation of the type (\ref{nrsusy}) 
must be accompanied by a compensating (non-relativistic) kappa-gauge 
transformation. This leads to transformations for which only $\delta\T_+$ 
and $\delta x^a$ are non-zero:
\be
\delta \T_+ = \ep_+ + {1\over2}\gamma^i\pa_i{x}^a\Gamma_a\ep_-, \qquad
\delta x^a = 2i \bar\T_+\Gamma^a\ep_-.
\ee
We see that $\T_+$  is the Goldstone fermion for a fermionic `shift' symmetry, 
with parameter  $\ep_+$, which is all that is left of the 
non-linearly realized supersymmetry of the gauge-fixed relativistic action. 
The parameter $\ep_-$ is the parameter of a linearly-realized worldvolume
supersymmetry.

The condition for supersymmetry preservation is $\delta\T_+=0$. As can be 
explicitly verified, this is the non-relativistic limit of the condition
$\Gamma_\kappa \epsilon = -\epsilon$. In the relativistic case, the 
supersymmetry preserving condition is $\Gamma_\kappa \epsilon = 
\pm\epsilon$, but the non-relativistic limit forces a choice of sign.


\section{The D0-brane}

The super D0-brane \cite{Bergshoeff:1996tu} is a direct 
generalization of the Lagrangian 
for the relativistic massive superparticle in four dimensions 
\cite{Casalbuoni:1976tz, deAzcarraga:1982dw}. In view of our 
discussion in the introduction  of the meaning of the non-relativistic 
limit, we set $c=1$.  The Lagrangian  is 
\be\label{Lag1}
L = -M\left(\sqrt{-\pi^2}\; -\;i\bar\T \Gamma_{11}\dot\T\right), \qquad 
\pi^m\;:=\;\dot X^m\;+\;{i}\bar\T \Gamma^m \dot\T.
\ee
The action is invariant under time-reparametrizations and  the 
$\kappa$-symmetry gauge  transformations, 
\be
\D\T = \frac12(1+ \frac{\pi_m\Gamma^m}{\sqrt{-\pi^2}} \Gamma_{11})\kappa,
\quad \quad 
\D X^m=-i\bar\T \Gamma^m\D\T.
\label{kapa}
\ee

In order to take the  non-relativistic  limit, we set 
\be
M=\omega\, m, \qquad X^0 = \omega\, x^0, \qquad 
\theta = \sqrt{\omega}\; \theta_- + {1\over\sqrt{\omega}}\, \theta_+
\ee
where $\theta_\pm$ are $\pm1$ eigenspinors of\footnote{This differs 
from  (\ref{pstar}) for the reason stated earlier.}
\be\label{dzerostar}
\Gamma_*= \Gamma^0\Gamma_{11}. 
\ee
Each has $16$ linearly-independent real components. We now have
\be
\pi^0 = \omega \, e \; +\;\frac{i}{{\w}}\; \bar\T_+\Gamma^0\dot\T_+,\qquad  
\pi^a =u^a 
\label{defua}
\ee
where
\be
e = \dot x^0+{i} \bar\T_- \Gamma^0 \dot\T_-\, \qquad
u^a = \dot x^a + 2i \bar\T_+\Gamma^a \dot \T_-, \qquad
x^a = X^a+i \bar\T_-\Gamma^a  \T_+,
\ee
and 
 \bea
-i\bar\T \Gamma_{11}\dot\T&=&
-i\omega\, \bar\T_-\Gamma^0\dot\T_- +
\frac{i}{\omega}\, \bar\T_+\Gamma^0 \dot\T_+.
\eea
The  \lag \bref{Lag1} can now be shown to take the form 
\be
L= -\omega^2 m\, \dot x^0 + {m\over 2 e} \, |{\bf u}|^2 - 2im\;
\bar\T_+\Gamma^0\dot\T_+ + {\cal O}\left(1/\omega^2\right). 
\ee
The first term, which is singular as $\w\rightarrow\infty$,  is the same 
total time derivative that we discarded in the purely bosonic case, and 
we can do the same again here;  note that this feature depends on a 
cancellation that occurs only for the kappa-symmetric action. The 
$\omega\rightarrow\infty$ limit may now be taken and we thus arrive 
at the time-reparametrization invariant form of the non-relativistic 
D0-brane Lagrangian
\be
L={m\over 2 e} \, |{\bf u}|^2 + 2im\; \T_+^T \dot\T_+ 
\label{Lag14}
\ee

Naturally, the action is still  invariant under supersymmetry, and 
$\kappa$-symmetry gauge transformations, albeit in modified form. 
In terms of the parameters $\ep_\pm$ we have
\be\label{susy1}
\D\T_\pm = \ep_\pm,\quad \qquad
\D x^0= i\bar\T_-\Gamma^0\ep_-,\quad \qquad
\D x^a\;=\;2 i\bar\T_-\Gamma^a\ep_+.
\ee
The non-relativistic kappa gauge transformations involve only $\kappa_-$ 
and are 
\bea
\D\T_-&=&\kappa_-,\quad\qquad\quad\,
\D\T_+\;= -\frac{1}{2e}\,\Gamma^0  u^a\Gamma_a \kappa_-
\nonumber \\
\D x^0&=&-i\bar\T_- \Gamma^0\kappa_-,\quad \;
\D x^a=-2{i}\bar\T_+ \Gamma^a\kappa_-. 
\label{covkap3}
\eea

We may fix the kappa gauge transformation by the gauge choice $\T_-=0$. 
If we also fix the time reparametrization invariance by choosing 
$\dot x^0=1$, then we find the fully gauge fixed Lagrangian 
\be
L =\frac{m}{2} |\dot {\bf x}|^2 \;+ \;2 {i}m (\T_+)^T \dot \T_+, 
\label{lgf}
\ee
which is quadratic in the $8$  real `bosons' ${\bf x}$ and the $16$ 
real `fermions' $\T_+$. The spacetime supersymmetry transformations 
of the variables of the 
Lagrangian must be accompanied by a compensating $\kappa$-transformations  
in order to maintain the choice of gauge. The resulting non-zero  
`physical'  supersymmetry transformations are
\be
\D\T_+= \ep_+ + \;\frac{1}{2}\Gamma^{0} {\dot x}^a\Gamma_a\,  
\ep_-\, ,\qquad\qquad
\D x^a=2{i}\bar\T_+ \Gamma^a\ep_-.
\ee

\vs

\section{Super-Bargmann algebra}

We have seen how the supersymmetric action for the non-relativistic D0-brane
emerges from the relativistic action in a suitable limit. Here we shall 
show that an analogous limit applied to the 10-dimensional $N=2$ 
super-Poincar\'e algebra yields the super-Bargmann symmetry algebra 
of the non-relativistic D0-brane. We should note here that other
supersymmetrizations of the Bargmann algebra have been considered in
other contexts. For example \cite{deAzcarraga:1982dw}; in addition, 
super-Bargmann algebras with vector supercharges appear in 
\cite{Gauntlett:1990xq,Duval:1993hs}. However, to the best of our knowledge,
the particular super-Bargmann algebra that we present here has 
not previously appeared in the literature.  

The non-zero (anti)commutation relations of the 10-dimensional $N=2$ 
super-Poincar\'e algebra, with central charge ${\cal Z}$, are 
(supressing spinor indices)
\bea
\left[P_m,~M_{pq}\right]&=&-i(\h_{mp}P_q-~\h_{mq}P_p), \nonumber
\\
\left[M_{mn},~M_{pq}\right]&=&
-i(\h_{np}M_{mq}~-~\h_{mp}M_{nq}~-~
\h_{nq}M_{mp}~+~\h_{mq}M_{np}), \nonumber
\\
\left[Q,~M_{mn}\right]&=& {i\over2}Q \Gamma_{mn}, \nonumber
\\
\{Q,~Q\}&=&2(C\Gamma^m) P_m ~+~ 2(C\G11){\cal Z}.
\label{QQ10}
\eea
For the purpose of taking the non-relativistic limit we set
\be\label{rescalingnr}
P_0= - {1\over2\w}\, H - \w Z, \qquad 
{\cal Z} =\omega Z- {1\over 2\w}\,  H,  \qquad M_{a0}=\w B_a. 
\ee
and
\be\label{rescalingnr2}
Q=  \sqrt{\w}\, Q_+ + {1\over\sqrt{\w}}\, Q_-\, ,
\ee
where $Q_\pm$ are $\pm1$ eigenspinors of the matrix $\Gamma_*$ of 
(\ref{dzerostar}). Note that
the rescaling of the components of $Q$ is `opposite' to the rescaling of 
the components of $\T$ in (\ref{newrescale}), so that $Q\T$ remains
finite as $\w\rightarrow\infty$. The super-Bargmann algebra is obtained in 
the $\w\rightarrow\infty$ limit. Using the identities
\be
\Gamma_{ab}\bP_\pm=\bP_\pm\Gamma_{ab}, \qquad 
\Gamma_{0a}\bP_\pm=\bP_\mp\Gamma_{0a}, \qquad 
\bP_\pm^T C = C\bP_\mp 
\ee
where
\be
\bP_\pm= {1\over2} \left(1\pm \Gamma_*\right) ,\qquad (\Gamma_*= \Gamma^0
\Gamma_{11}), 
\ee
one finds that the non-zero (anti)commutators in the 
$\w\rightarrow\infty$ limit are 
\bea
\left[M_{ab},M_{cd}\right] &=&-i\left(\eta_{bc}M_{ad} - 
\eta_{ac}M_{bd} - \eta_{bd}M_{ac}+ \eta_{ad}M_{bc}\right) \nn\\
\left[P_a,M_{bc}\right]  &=& -i\left(\eta_{ab}P_c - \eta_{ac}P_b\right), \nn\\
\left[B_a,M_{bc}\right] &=& -i\left(\eta_{ab}B_c - \eta_{ac}B_b\right), \nn\\
\left[H,B_a\right]  &=&  iP_a \nn\\
\left[Q_\pm, M_{ab}\right] &=& {i\over2}Q_\pm\Gamma_{ab} \nn\\
\left[Q_-,B_a\right] &=& {i\over2}Q_+\Gamma_{a0} \nn\\
\left\{Q_-,Q_-\right\} &=& 2H\bP_- \nn\\
\left\{Q_+,Q_-\right\} &=& 2\left(C\Gamma^a\bP_-\right) P_a 
\eea
which defines the super-Galilei algebra, and
\be\label{centralext}
\left[P_a,B_b\right]= i\delta_{ab} Z \qquad  
\left\{Q_+,Q_+\right\} = 4Z\bP_+ ,
\ee
which extends the super-Galilei algebra to the super-Bargmann 
algebra via the central charge $Z$.
Note that the sub-superalgebra spanned by $(H,Q_-)$ 
is a conventional $N=16$ worldline supersymmetry algebra.

The super-Bargmann algebra is realized as a  symmetry algebra 
of the non-relativistic D0-brane \lag \bref{Lag14}. The even generators are 
\bea
H &=&-p_0,\qquad P_a~=~p_a, \qquad
J_{ab} = x_{[a}p_{b]}-{1\over2}\z_+~\Gamma_{ab}~\T_+
-{1\over2}\z_-~\Gamma_{ab}~\T_-,
\nonumber \\
B_a &=& p_a x^0-{1\over2}\z_+~\Gamma_{a0}~\T_- 
-m\left(x_a+~i\Tb_+\Gamma_a\T_-\right) -
\frac{i}{2}\Tb_-\Gamma_{ba}\Gamma_0\T_-p^b,
\nn\\
Z&=&m
\eea
and the odd generators are
\be
 Q_-=-i~\z_-+\Tb_-(\Gamma^0p_0), \qquad 
Q_+=-i~\z_++2\Tb_-(\Gamma^ap_a)-2m~\Tb_+\Gamma^0.
\ee
Note that the central charge $Z$ is the particle's mass. Both occurrences 
of it in (\ref{centralext}) 
arise from the central charge ${\cal Z}$ in the original 
super-Poincar\'e algebra. 
The non-relativistic particle and the relativistic 
massive superparticle have both frequently been  used to illustrate the 
relevance of central charges in particle mechanics (see, for example, 
\cite{Gauntlett:1990nk}). It is thus amusing to note (in agreement with 
a result of  \cite{deAzcarraga:1991fa} in a different context) that these 
two examples become essentially the same example in the context of the 
non-relativistic D0-brane! 

To conclude this section, we observe that there is a 
one-parameter family of contractions of the Poincar\'e algebra, 
obtained by setting
\be
P_0=-{1\over2\w} H - \w Z, \qquad 
{\cal Z} =k\omega Z- {1\over2\w} H,  \qquad M_{a0}=\w B_a
\ee
for constant $k$. The  $k=1$ case was considered above. In general,
 one finds (as before)
the Bargmann algebra with central charge $Z$ in
the $\w\rightarrow\infty$ limit. For any $k$ there is a contraction of the
super-Poincar\'e algebra obtainable by setting
\be
Q= \sqrt{\w} S. 
\label{nnel}
\ee
This leads to the direct sum of the Bargmann algebra with a 
32-dimensional odd algebra defined by the anticommutator
\be
\{{S},{S}\} = 2 Z (1 + k \Gamma^0\Gamma_{11}). 
\ee
This is not a supersymmetry algebra, but it {\it is} the 
superalgebra realized by the $p=0$ analog of the Lagrangian 
(\ref{freefield}) for $k=Q/T$ (i.e., a non-BPS D0-brane for $k\ne1$).   
For $|k|=1$, {\it but not otherwise}, the alternative rescaling 
(\ref{rescalingnr2}) of the fermionic generators is possible. 
As we have seen this
alternative rescaling allows a contraction to a 
non-relativistic superalgebra that contains a standard 
worldline supersymmetry algebra as a sub-superalgebra.

\section{The IIA superstring}

\vs

We now turn to the IIA superstring. The action is \cite{Green:1983wt}
\be
S= -T  \int d^2\sigma \, \left[ \sqrt{-\det G} -  {\cal L}_{WZ}\right]
\label{LGS}
\ee
where the WZ Lagrangian density is
\be
{\cal L}_{WZ}\;=\;-\vep^{jk}{i}\, \Tb\Gamma_m\Gamma_{11}\pa_j\T\; 
(\Pi_k{}^m -\frac{ i}{2}\Tb\Gamma^m\pa_k\T)
\ee
The action is invariant under the kappa-gauge transformation
\be\label{kappaGS}
\D\T = \frac12(1-\Gamma_{\kappa})\kappa,\quad 
\D X^m = -{i}\Tb \Gamma^m\D\T \
\ee
where
\be
\Gamma_{\kappa}\;\equiv\;\frac12\frac{\vep^{jk}}{\sqrt{-G}}
(\Pi^m_j\Gamma_m)(\Pi^n_k\Gamma_n)\G11. 
\ee

The non-relativistic limit for strings is special in that
 {\it there is no need to rescale the string tension}.
Only the dynamical variables need be rescaled, and we do this by setting
\bea
X^\mu = \omega\, x^\mu \, , \quad \T\;=\;{\sqrt{\w}}\;\T_-~+~
\frac{1}{\sqrt{\w}}\;\T_+, 
\label{contraction}
\eea
where $\T_\pm$ are the $\pm1$ eigenspinors of the matrix 
\be
\Gamma_*= \Gamma_0\Gamma_1\Gamma_{11}. 
\ee
We now have
\bea
\Pi_i{}^\mu&=&\w\; e_i{}^\mu +\frac{i}{{\w}}\;\Tb_+\Gamma^\mu\pa_i\T_+,
\quad \quad \quad 
\Pi_i{}^a~=~u_i{}^a. 
\label{defua2}
\eea
where 
\be
e_i{}^\mu = \partial_i x^\mu +{i}\Tb_-\Gamma^\mu \partial_i\T_-, 
\ee
and
\be
u_j{}^b ~=~\pa_j x^b+{2i}\Tb_+\Gamma^b\pa_j\T_-, \qquad 
\left(x^a~=~X^a+{i}\Tb_-\Gamma^a\T_+\right).
\ee

Proceeding as before, we find that the terms in the action proportional to 
$\w^2$ combine to form a term proportional to 
$\vep^{jk}\vep_{\mu\nu}(\pa_j x^\mu\pa_k x^\nu)$. This is a total derivative,
 which we may drop. Taking the $\omega\rightarrow\infty$ limit then 
yields
\bea
\CL &=&  -{1\over2} e\,  \cg^{jk}\;{\bf u}_j \cdot {\bf u}_k \;-\;2i\;
e\,(\Tb_+\cga^i\pa_i\T_+) \nonumber\\
&& - 2i\;\vep^{jk}
\left(\Tb_+\Gamma_a\Gamma_{11}\;\pa_j\T_-\right)
\left(u_k{}^a -{i}\;\Tb_+\Gamma^a\;\pa_k\T_-\right).
\label{Lag22}
\eea
The spacetime supersymmetry transformations are obtained in the way explained
earlier, with the result that
\be\label{covsusy}
\D\T_\pm = \ep_\pm ,\quad  
\D x^\mu = {i}\Tb_-\Gamma^\mu\ep_-,\quad 
\D x^a\;= 2{i}\Tb_-\Gamma^\mu\ep_+ .
\ee
Using the invariance of the 1-forms $e^\mu, \ {\bf u} $ and $d\T$, 
and the standard cyclic identity of the 10-dimensional Dirac matrices, 
one can show that the variation of the Lagrangian \bref{Lag22} is 
a total derivative. 
\vs
The non-relativistic kappa gauge transformation is also found  as 
before, with the result that
\bea\label{covkappa}
\D\T_-  &=&  \kappa_-,\quad \qquad \quad
\D\T_+= -\frac{1}{2}\cga^i {u_i}^a\Gamma_a \kappa_- \nonumber\\
\D x^\mu& =& -{i}\Tb_-\Gamma^\mu\kappa_-,\quad 
\D x^a\;=\;-2i\Tb_+\Gamma^a\kappa_-.
\eea
This gauge invariance allows us to choose  $\T_-=0$. If we further 
fix the reparametrization 
invariance by the usual Monge gauge then we arrive at the 
non-relativistic IIA superstring action
\be
S= -T\;\int d^2\xi \left[ \frac{1}2\;\h^{ij} \pa_i  
{\bf x} \cdot \pa_j {\bf x} \;+\;2
i\; \Tb_+\Gamma^i\pa_i\T_+\right]
\label{Lag22gf2}
\ee
The supersymmetry transformations of the gauge-fixed variables 
appearing in this action  
are obtained by combining the supersymmetry transformation 
with a compensating kappa-transformation, with the result that
\be
\D\T_+= \ep_+\;+\;\frac{1}{2}\gamma^i\pa_i x^a 
\Gamma_a \ep_-,\quad \quad 
\D x^a\;=\;2i\Tb_+\Gamma^a\ep_-.
\label{ressusyst}
\ee

\section{Conclusions and Comments}

In this paper we have shown (extending earlier work 
\cite{Townsend:1999hi, Gomis:2000bd, Danielsson:2000gi, Garcia:2002fa})
that there is a natural p-brane generalization of the non-relativistic 
limit of the action for a point particle in which the limiting 
gauge-fixed action is a worldvolume Poincar\'e invariant field 
theory. The main result of this paper is a generalization of this
non-relativistic limit to superbranes. We have shown that there
is a limit of the reparametrization invariant and kappa-gauge invariant
super-p-brane (made possible by a crucial cancellation
between the Dirac and WZ Lagrangians of terms which are separately
singular in the limit)  with  the feature that the spacetime
supersymmetry of the
gauge-invariant action implies a worldvolume supersymmetry of the
gauge-fixed action, exactly as happens for relativistic superbranes.
An interesting feature of this limit is that the non-relativistic 
kappa-gauge invariance is irreducible, in contrast to the reducibility
of the relativistic kappa-gauge invariance. 
Our results apply to all cases in which the supersymmetric worldvolume 
field theory involves only scalars and spinors, and we have given explicit
Lagrangians for $p=0,1,2$, including the gauge-invariant Lagrangian 
of the non-relativistic M2-brane. We expect that a 
similar result will apply to  super-D-p-branes and the M5-brane.

We have discussed the case of the D0-brane in  detail. In particular, 
we have shown that our non-relativistic limit corresponds to a 
particular contraction of the super-Poincar\'e symmetry algebra 
of the relativistic D0-brane. 
The contracted algebra is a supersymmetric extension of the Bargmann 
algebra, which is indeed realized by the non-relativistic D0-brane, 
and which contains a conventional $N=16$ worldline supersymmetry 
algebra as a sub-super-algebra. One can anticipate that similar 
results will apply for $p>0$, thus  extending to superbranes the 
non-relativistic brane algebras introduced in \cite{Brugues:2004an}. 

We have also discussed in detail the case of the IIA superstring, obtaining 
the reparametrization invariant and kappa-symmetric action for a 
non-relativistic IIA superstring.  We were motivated to consider 
this case separately  because explicit results in this case could 
be useful for a supersymmetric  extension of the non-relativistic 
string theory of  \cite{Gomis:2000bd, Danielsson:2000gi}. 

Our treatment of the non-relativistic limit of p-branes for general
 $p$  brings to light a special 
feature of the $p=1$ case. Recall that the limit generally involves 
a rescaling of the p-brane tension. For example, for $p=0$ the limit
involves a rescaling in which the ratio $\omega= M/m$ of the 
relativistic particle mass $M$ to the non-relativistic particle 
mass $m$ is taken to infinity. In other words, for fixed $m$ 
one takes $M\rightarrow\infty$, as expected (given that we 
have fixed the speed of light) because this corresponds to 
an effectively infinite rest-mass energy that makes particle-antiparticle 
pair production impossible. Curiously, for $p=2$ one has the 
inverse relation $\omega= T_{nr}/T_{rel}$ for non-relativistic 
and relativistic tensions, so that the non-relativistic limit 
corresponds to taking $T_{rel}$ to zero for fixed $T_{nr}$ 
(and fixed speed of light). This can be understood by viewing the 
relativistic p-brane action as a $(p+1)$-dimensional field theory 
with coupling constant $g\sim 1/\sqrt{T_p}$. For $p\ge 1$ the non-relativistic
limit is an ultra-violet (UV) limit because we focus on a small 
patch of the brane and then rescale so that the patch  
becomes the entire brane. For $p\ge 2$ the coupling constant 
$g$ is dimensionful and is driven to infinity in the UV limit, whereas
for $p=1$ the coupling constant is dimensionless and  
no (classical) scaling is needed. 

Finally, we should emphasise that we have concentrated in 
this paper on the non-relativistic limit
of superbranes in a Minkowski vacuum. There are further interesting 
issues in relation to non-flat backgrounds that we hope to 
address in future work. For example, the ideas
developed here may have implications for the `extra-supersymmetry'
puzzle arising in the field theory limit of the M2-brane in certain
hyper-K\"ahler backgrounds \cite{Townsend:2002wd}.

\vskip 1cm
\noindent
{\bf Acknowledgements:} 
KK and PKT thank the University of Barcelona for hospitality.
 PKT thanks ICREA for financial support.
 We are grateful to Jaume Gomis and Machiko Hatsuda for discussions.
 This work is partially supported by the European Commission RTN
 program under contract 
HPNR-CT-2000-00131 and by MCYT FPA 2001-3598 and CIRIT GC 2001SGR-00065.

\end{document}